\documentclass[prl,twocolumn,showpacs,amssymb,floatfix]{revtex4}

\usepackage{epsfig}
\begin{document}
%\draft

%\twocolumn
 
%\narrowtext
\title{Roughness of a Tilted Anharmonic String at Depinning}

\author{T. Goodman$^a$ and S. Teitel$^b$}
\affiliation{$^a$Department of Electrical Engineering, Bucknell University, Lewisburg, PA 17837\\$^b$Department of Physics and Astronomy,  University of Rochester, Rochester, NY 14627}
\date{\today} 
\begin{abstract}
We consider the discretized model of a driven string with an anharmonic elastic energy, in a two dimensional random potential, as introduced by Rosso and Krauth \protect\cite{R1}.
Using finite size scaling, we numerically compute the roughness of the string in a uniform
applied force at the critical depinning threshold.  By considering a string with a net
average tilt, we demonstrate that the anharmonic elastic energy crosses the model over
to the quenched KPZ universality class, in agreement with recent theoretical predictions.

\end{abstract}
\maketitle

Recently, Rosso and Krauth (RK) reported \cite{R1} new 
simulations of roughening at the depinning 
threshold of a driven one dimensional string in a two 
dimensional (2d) random potential.  Introducing higher order 
anharmonic terms to a quadratic elastic energy for the string, RK 
found a value for the roughness exponent $\zeta\simeq 0.63$,
in contrast to the values $\zeta\sim 1.2$ found in earlier 
simulations \cite{X1}, and recently theoretically \cite{X2},
using a purely quadratic energy.  RK noted that the
value $\zeta\simeq 0.63$ had previously been found in
some cellular automata models for depinning \cite{X3}.

In a subsequent work \cite{R2} RK (with Hartmann)
noted that that when an average tilt is applied to the 
string, the anharmonic terms break the 
rotational invariance present in the quadratic model, thus
suggesting that the anharmonic terms might cross the model
into the quenched KPZ universality class, previously
introduced by Kardar \cite{R3} to explain the {\it anisotropic} depinning 
observed in the automata models.  Simulations \cite{R4} of
a continuum model with the quenched KPZ term
found $\zeta\simeq 0.61\pm 0.06$, consistent with the automata models
and with the anharmonic model of RK.  Most recently, a functional renormalization
group calculation by Le Doussal and Wiese \cite{R5} argued that
the quenched KPZ term can indeed be generated, not only by the anisotropic
disorder considered by Kardar, but also by the anharmonic elastic energy terms
introduced by RK.

A key prediction of Kardar's for anisotropic depinning is that the 
roughness exponent for a {\it tilted} interface will differ from that of 
an untilted one; for a tilted string in 2d, he predicted the exact 
value of $\zeta_{\rm tilt}=1/2$.
In this report, by computing the $\zeta_{\rm tilt}$ of RK's model for 
the first time, we offer a direct 
numerical demonstration that their model does
indeed belong in the quenched KPZ universality class.

Our model is the same as that of RK.  We take for the energy of the string,
\begin{equation}
E[h_i]=\sum_{i=0}^{L-1}\left\{V(i,h_i)-fh_i+E_{\rm 
el}(h_{i+1}-h_i)\right\}\enspace,
\label{e1}
\end{equation}
where $h_i$, is the integer height of the string at position $i=0,\ldots L$ 
on a discretized lattice, 
$V(i,j)$ is an uncorrelated random Gaussian potential with zero average and  unit 
variance, $f$ is a uniform external driving force, and $E_{\rm el}$
is the elastic energy of deforming the string.  $V(i,j)$ 
is taken periodic on an $L\times L$ system size.  In their work,
RK used periodic boundary conditions.  Here, to
model a tilted interface with net slope $s$, we use boundary 
conditions $h_{L}=h_{0}+sL$.  Defining the height relative
to a uniformly tilted line, $\delta 
h_{i}\equiv h_{i}-si$, so that $\delta h_{L}=\delta h_0$,
we can rewrite Eq.(\ref{e1}) in terms of 
the $\delta h_i$ and recover the same model as RK
except that the elastic term now has the form,
\begin{equation}
\sum_{i=0}^{L-1}E_{\rm el}(\delta h_{i+1}-\delta h_{i}+s)\enspace.
\label{e2}
\end{equation}

We now carry out simulations of 
Eq.(\ref{e1}), using the elastic term of Eq.(\ref{e2}).  Using the 
same algorithm \cite{R6} as RK, we consider slopes $s=0$ and $s=1$
for one of the specific cases studied in \cite{R1}, 
\begin{equation}
E_{\rm el}(\Delta)=\Delta^{4}/16\enspace.
\label{e3}
\end{equation}
We compute the interface roughness $W$ for a system of length $L$,
\begin{equation}
W^2\equiv{1\over L}\sum_{i=0}^{L-1} \left[(\delta h_i^c-\overline{\delta h^c})^2\right]\sim L^{2\zeta}\enspace,
\label{e4}
\end{equation}
where $\delta h_i^c$ is the relative height at site $i$ of the critical string at depinning,
$\overline{\delta h^c}$ is the average relative height of the critical string, $[\dots]$ represents
an average over many realizations of the random potential $V(i,j)$, and $\zeta$
is the roughness exponent.  We
also compute the disorder average of the critical force, $f_c$.

\begin{figure}[h]
\epsfxsize=7.5truecm
\epsfbox{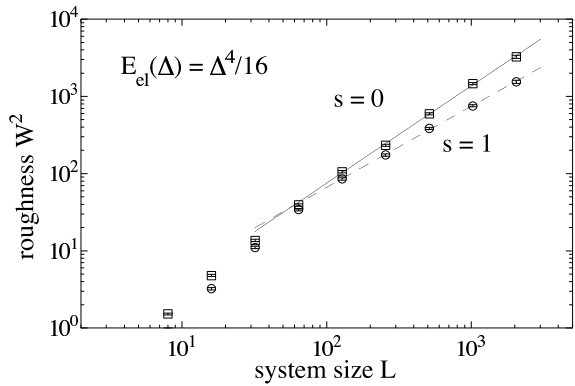}
\caption{Roughness $W^2$ vs. system size $L$, for strings of net 
slope $s=0$ and $s=1$. The lines are the best fits to $W^2\sim L^{2\zeta}$
using system sizes $L=128 - 2048$.
}
\label{f1}
\end{figure}

Our results for string roughness $W^{2}$ vs. $L$, averaged over $500$
disorder samples (for $L=2048$ we use only $200$ samples), are plotted in Fig.~\ref{f1}.
For $s=0$, our numerical values agree with those in Ref.\cite{R1}.
The straight lines on the log-log plot indicate the power law relation, $W^2\sim L^{2\zeta}$,
and the difference in slopes indicate clearly different roughness exponents for the
tilted ($s=1$) and untilted ($s=0$) strings.

\begin{figure}
\epsfxsize=7.5truecm
\epsfbox{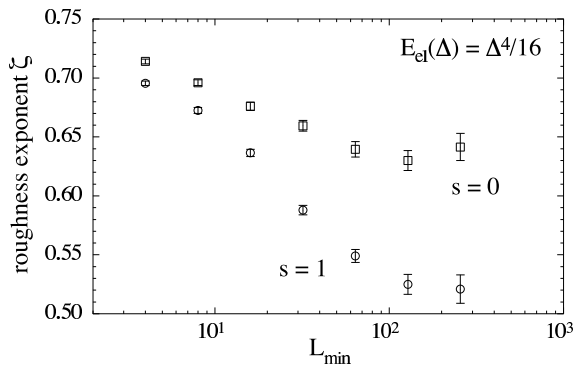}
\caption{Roughness exponents $\zeta$ for the tilted ($s=1$) and untilted ($s=0$)
strings, as obtained by fitting the data of Fig.~\protect\ref{f1} to $W^2\sim L^{2\zeta}$
using data for sizes $L_{\rm min}$ to $20484$.
}
\label{f2}
\end{figure}

To determine the values of the
exponent $\zeta$, we fit the data in Fig.~\ref{f1} to $W^2\sim L^{2\zeta}$, using
system sizes from $L_{\rm min}$ to $L=2048$.  We plot the resulting values of $\zeta$ 
vs. $L_{\rm min}$ in Fig.~\ref{f2}, for $L_{\rm min}=4$ to $256$.  We see that
as $L_{\min}$ increases, the values of $\zeta$ decrease and saturate to a constant value
characterizing the roughness in the asymptotic large $L$ limit.  Using the results
from fitting with $L_{\rm min}=128$ we find for $s=0$ the value $\zeta\simeq 0.63\pm 0.01$.
This value is used to plot the solid straight line in Fig.~\ref{f1}, and agrees with the
value found by RK.  For $s=1$, however, we find the value $\zeta=0.52\pm 0.01$.
We use this value to plot the dashed line in Fig.~\ref{f1}.  Given that $\zeta$ for $s=1$
still shows a small systematic decreases as $L_{\rm min}$ increases, we believe
our value is in excellent agreement with the exact value of $1/2$ predicted
by Kardar, and thus verifies that the anharmonic model of RK is in the
quenched KPZ universality class.

\begin{figure}
\epsfxsize=7.5truecm
\epsfbox{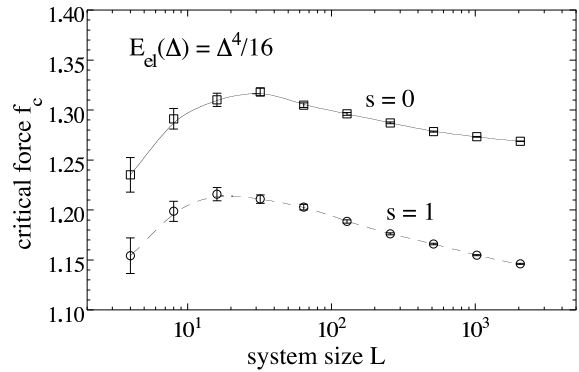}
\caption{Critical force $f_c$ vs. system size $L$, for strings of net 
slope $s=0$ and $s=1$.  The lines are guides to the eye.
}
\label{f3}
\end{figure}

Finally, in Fig.~\ref{f3} we plot the critical force $f_c$ as a function of
system size $L$ for the tilted and untilted strings.  We see clearly that the
critical forces approach different values as $L$ increases, another signature
of anisotropic depinning.

We thank T. Nattermann and Y. Shapir for helpful discussions.
This work has been supported by DOE grant DE-FG02-89ER14017 and 
NSF grant PHY-0242483.

\end{document}